\documentclass{ws-ijmpd}
\usepackage{graphicx}

\begin{document}

%

\def\nocropmarks{\vskip5pt\phantom{cropmarks}}


%


%
\catchline{}{}{}
%

\title{Microlensing signature of a white dwarf population in
the galactic halo}

\author{M. HAFIZI}

\address{Department of Physics, University of Tirana, Albania}

\author{F. DE PAOLIS, G. INGROSSO, A.A. NUCITA}

\address{Department of Physics and INFN, University of Lecce, CP 193,
I-73100 Lecce, Italy}

\maketitle

\pub{Received (received date)}{Revised (revised date)}

\begin{abstract}
Microlensing and pixel-lensing surveys play a fundamental role in
the searches for galactic dark matter and in the study of the
galactic structure. Recent observations suggest the presence of a
population of old white dwarfs with high proper motion, probably
in the galactic halo, with local mass density in the range
$1.3\times 10^{-4}-4.4\times 10^{-3}~M_{\odot}$ pc$^{-3}$, in
addition to the standard galactic stellar disk and dark halo
components. Investigation of the signatures on microlensing
results towards the LMC of these different lens populations, with
particular emphasis to white dwarfs, is the main purpose of the
present paper. This is done by evaluating optical depth and
microlensing rate of the various lens populations and then
calculating through a Montecarlo program, the probability that a
lens which has caused a microlensing event of duration $t_{\rm E}$
belongs to a certain galactic population. Data obtained by the
MACHO Collaboration allow us to set an upper bound of $1.6\times
10^{-3}~M_{\odot}$ pc$^{-3}$ to the local mass density of white
dwarfs distributed in spheroidal models, while for white dwarfs in
disk models all values for the local mass density are in agreement
with observational results.
\end{abstract}

\section{Introduction}
MACHOs (Massive Astrophysical Compact Halo Objects) have been
detected since 1993 in microlensing experiments towards the Large
and Small Magellanic Clouds (LMC and SMC) as well as towards the
galactic center. Although the events detected towards the SMC seem
to be a self-lensing phenomenon \cite{sahu,ss,st,gyuk}, a similar
interpretation of all the events discovered towards the LMC looks
unlikely \cite{alcock2000}. Indeed, no purely LMC self-lensing
models may produce optical depths as large as that reported by
MACHO and EROS Collaborations. However, due to large uncertainties
and the few events at disposal, it is not possible to draw at
present sharp conclusions about the lens nature and localization.
Indeed, the most plausible solution is that the LMC events
detected so far are not due to a single lens population but to
lenses belonging to different intervening populations: low mass
stars in the LMC or in the galactic disk and MACHOs in the
galactic halo. The average MACHO mass from observations results to
be $\simeq 0.5~M_{\odot}$, \cite{jms}, so that the white dwarf
hypothesis looks to several authors as the best explanation for
MACHOs. However, the resulting excessive metallicity of the halo
makes this option untenable, unless the white dwarf contribution
to halo dark matter is not substantial. \cite{gm,binney} So, some
variations on the theme of white dwarfs have been explored.

An option is that the galactic halo resembles more closely a
minimal halo (i.e. the Milky Way is well described by a maximal
disk model) rather than an isothermal sphere, in which case the
average MACHO mass gets decreased so that most of them may still
be brown dwarfs. In this connection, two points should be
stressed. First, a large fraction of microlensing events
(constituting up to $50\%$ in mass) can be binary systems - much
like ordinary stars - thereby counting as twice more massive
objects. \cite{depaolismnras,distefano} Second, MACHOs in the
galactic halo could be brown dwarfs - with mass substantially
larger than $\simeq 0.1~M_{\odot}$ \cite{HANSEN} - since a slow
accretion mechanism from cloud gas is likely to occur. \cite{lcs}
An alternative possibility has been pointed out: since the stellar
initial mass function may change with the galactocentric distance,
\cite{taylor} it can well happen that brown dwarfs substantially
contribute to the halo mass density without however dominating the
microlensing optical depth. \cite{ke}

On the other hand, it has been suggested that, in addition to main
sequence stars, brown dwarfs and white dwarfs, also a population
of black holes may exist in the Galaxy. These black holes may have
been already observed in microlensing surveys.
\cite{bennett,quinn} Indeed, at least 6 extremely long events
exhibiting very strong microlensing parallax signals have been
detected by MACHO, GMAN and MPS Collaborations, indicating that a
substantial fraction of the galactic lenses can be massive stellar
remnants with masses up to $\simeq 10~M_{\odot}$.

Quite recently, faint blue objects discovered by the Hubble Space
Telescope have been understood as old halo white dwarfs lying
closer than $\sim 2$ kpc from the Sun:
\cite{hansen,hansen01,ibata} they look as a good candidate for
MACHOs. Moreover, very recently it has been found that the sample
of local white dwarfs is largely complete out to 13 pc and that
the local number density of white dwarf stars is $n_{\rm
wd}(R_0)=(5.0 \pm 0.7) \times 10^{-3}$ pc$^{-3}$. \cite{holberg}

If these white dwarfs make a relatively important fraction of the
galactic dark matter, as it seems, then they should also show up
in the microlensing searches. In this context, we consider either
a spheroid or a disk-like distribution of white dwarfs in addition
to the two (thin and thick) stellar disks, the bulge and the
standard halo components. \cite{bss} Investigation of the
signatures on microlensing results towards the LMC of the
different lens populations, with particular emphasis to white
dwarfs, is the main aim of the present paper.

To this end, we shall consider all available informations from
gravitational microlensing experiments towards the LMC: optical
depth, microlensing rate and event duration to try solving the
problem of distinguishing among different lens populations. Note
that we do not take into account the informations from the two
observed parallax events towards the SMC which imply that these
events are most likely self-lensing. From this fact one cannot
infer that also all LMC events are self-lensing events since it is
known that the SMC is extended along the line of sight (so that
microlensing is dominated by self-lensing)  while there is little
evidence that the LMC is similarly extended. Furthermore, since we
are interested in the matter distribution in our galaxy and are
not modelling the LMC mass distribution, we do not calculate the
expected number of self-lensing events but use the estimation by
\cite{jms} to subtract these events from the total number of
microlensing events given by the MACHO Collaboration.
\cite{alcock2000}

Regarding the problem of distinguishing between different lens
populations, we also note that since white dwarfs lie relatively
nearby the stellar disk, their transverse velocity should be
different with respect to the typical transverse velocity of the
halo MACHOs. This should give, in principle, a ``signature'' of
the events due to white dwarfs in microlensing searches.

In Section 2 we review the current status of high velocity white
dwarf observations from which the local white dwarf mass density
value is estimated. In Section 3 we present the adopted mass
distribution for the stellar, white dwarf and halo components
while in Section 4 we briefly review the standard equations for
the microlensing optical depth, event rate and event duration. In
Section 5 we present our model results which, compared with
microlensing observations towards the LMC, allow to constrain the
white dwarf local mass density. Finally, in Section 6 we draw our
main conclusions.

\section{The ``halo'' white dwarf component}

Recently the detection of a significant population of old white
dwarfs with high proper motion which might be representative of
the galactic halo has been announced. \cite{oppenheimer} Assuming
white dwarfs with mass $m_{\rm wd}\simeq 0.6~M_{\odot}$, the
inferred mass density  is $\rho_{\rm wd}(R_0)=1.3\times 10^{-4}~
M_{\odot}$ pc$^{-3}$. \footnote{This was derived by using the
$\frac{1}{V_{\rm max}}$ technique. \cite{wo}} This estimate (which
accounts only for $2\%$ of the local dynamical mass density) has
to be considered as a lower limit since a larger population of
even fainter and cooler white dwarfs may be present in the
galactic halo. \cite{oppenheimer} Moreover, faint blue objects
discovered by the Hubble Space Telescope have been understood as
old halo white dwarfs lying closer than $\sim 2$ kpc from the Sun.
\cite{hansen,ibata} More recently, it has been also found that the
sample of local white dwarfs is largely complete out to 13 pc and
that the local number density of white dwarf stars is $n_{\rm
wd}(R_0)=(5.0 \pm 0.7) \times 10^{-3}$ pc$^{-3}$ with a
corresponding mass density of $\rho_{\rm wd}(R_0)=(3.4 \pm 0.5)
\times 10^{-3}~ M_{\odot}$ pc$^{-3}$ for an assumed white dwarf
mass $m_{\rm wd}\simeq 0.65~M_{\odot}$. \cite{holberg}
\footnote{For comparison we note that the expected white dwarf
contribution from the standard stellar spheroid is only $\rho_{\rm
wd}(R_0)=1.3\times 10^{-5}~ M_{\odot}$ pc$^{-3}$. \cite{gfb}}

Therefore, if this white dwarf population is representative of the
galactic halo or belongs to a thick disk - as implied by the
analysis in \cite{rsh} - then it should obviously contribute to
the claimed microlensing populations. For completeness we mention
that alternative explanations have been suggested. For instance it
has been shown that the old white dwarf population might still be
interpreted as a high-velocity tail of the disk population
\footnote{However, it has been shown that the white dwarf sample
considered by Reid, Havley and Gizis \cite{rhg} seems to contain a
too large fraction of high velocity objects. \cite{hansen01}}.
\cite{rhg} We also note that this population of high-velocity
white dwarfs can be derived from a population of binaries residing
initially within the thin disk of the Galaxy. The binaries with a
massive enough star are broken up if the primary star explodes as
a Type II Supernova owing to the combined effects of the mass loss
from the primary and the kick received by the neutron star on its
formation. It has been shown that for a reasonable set of
assumptions concerning the galactic supernova rate and the binary
population, the obtained local number density of high-velocity
white dwarfs is compatible with that inferred from observations.
\cite{dkr} Therefore, a population of white dwarfs originating in
the thin disk may make a significant contribution to the observed
population of high-velocity white dwarfs. \footnote{More recently,
it has been presented an analysis of halo white dwarf candidates,
based on model atmosphere fits to the observed energy
distribution. Indeed, a subset of the high velocity white dwarf
candidates which are likely too young to be members of the
galactic halo was identified, thus suggesting that some white
dwarfs born in the disk may have acquired high velocities.
\cite{bergeron}}

In Section 5 and 6 we shall consider white dwarfs distributed in
the Galaxy, either with a spheroidal or disk-like shape, up to
very large distances. On the other hand, since only local
measurements for the average white dwarf mass density are
available, we shall consider the white dwarf average local mass
density as a free parameter constrained by the available
observations with lower bound $\rho_{\rm wd}(R_0)=1.3\times
10^{-4}~ M_{\odot}$ pc$^{-3}$ \cite{oppenheimer} and upper bound
$\rho_{\rm wd}(R_0)=4.4\times 10^{-3}~ M_{\odot}$ pc$^{-3}$.
\cite{gfb} We also use as a reference value the white dwarf mass
$m_{\rm wd}=0.65~M_{\odot}$.

\section{Mass distribution in the Galaxy}

We consider a four component model for the mass distribution in
the Galaxy: a triaxial bulge, a double stellar disk, a white dwarf
component (either with a disk-like or a spheroidal shape) and a
dark matter halo.

In particular, the central concentration of stars is described by
a triaxial bulge model with mass density given by
\begin{equation}
\rho_{\rm b}(x,y,z) =\frac{M_{\rm b}}{8\pi \tilde
abc}e^{-s^2/2}~,~~~{\rm with}~~~ s^4=(x^2/\tilde
a^2+y^2/b^2)^2+z^4/c^4~, \label{eq:2.1}
\end{equation}
where the bulge mass is $M_{\rm b} \sim 2 \times
10^{10}~M_{\odot}$ and the scale lengths are $\tilde a=1.49$ kpc,
$b=0.58$ kpc, $c=0.40$ kpc. \cite{dwek} The coordinates $x$ and
$y$ span the galactic disk plane, whereas  $z$ is perpendicular to
it. The remaining stellar component can be described with a double
exponential disk, \cite{gwk} so that the galactic disk has both a
``thin'' and a ``thick'' component. For the ``thin'' luminous disk
we adopt the following density distribution
\begin{equation}
\rho_{\rm d}(X,z) = \frac {\Sigma_0 } {2H}
~e^{-|z|/H}~e^{-(R-R_0)/h}, \label{eq:2.2}
\end{equation}
where the local projected mass density is $\Sigma_0 \sim
25~M_{\odot}$ pc$^{-2}$, the scale parameters are $H\sim 0.30$ kpc
and $h\sim 3.5$ kpc and $R_0$=8.5 kpc is the local galactocentric
distance. Here $R$ is the galactocentric distance in the galactic
plane. For the ``thick'' component we consider the same density
law as in eq. (\ref{eq:2.2}), but with variable thicknesses in the
range $H=1 \pm 0.5$ kpc and local projected density $\Sigma_0 \sim
35 \pm 15~M_{\odot}$ pc$^{-2}$.

For the halo component we consider a standard spherical halo model
with mass density given by
\begin{equation}
\rho_{\rm h}(r) = \rho_{\rm h}(R_0)\frac{a_{\rm h}^2
+R_0^2}{a_{\rm h}^2+r^2},
\end{equation}
where $a_{\rm h}\simeq 5.6$ kpc is the halo dark matter core
radius, $\rho_{\rm h}(R_0)$ is the local halo dark matter density.
As required by microlensing observations \cite{alcock2000} we
take, for definiteness, halo MACHOs of mean mass $m_{\rm h}\simeq
0.5~M_{\odot}$.

As far as the white dwarf component is concerned, we assume white
dwarfs of the same mass $m_{\rm wd}\simeq 0.65~M_{\odot}$,
distributed
according to two different laws: \\
a) a thick disk distribution with
\begin{equation}
\rho_{\rm wd}(R,z) = \frac{\Sigma_{\rm wd}(R_0)}{2H_{\rm wd}}
~e^{-|z|/H_{\rm wd}}~e^{-(R-R_0)/h}, \label{wddisk}
\end{equation}
or b) a spheroidal shape with
\begin{equation}
\rho_{\rm wd}(x,y,z) = \frac{\rho_{\rm wd}(R_0)}{q_{\rm wd}}
\frac{a_{\rm wd}^2+R_0^2}{a_{\rm wd}^2+x^2+y^2+z^2/q_{\rm wd}^2}~,
\label{wdhalo}
\end{equation}
where $a_{\rm wd}=2-4$ kpc is the white dwarf core radius, $q_{\rm
wd}$ is the flatness parameter and $H_{\rm wd}=1-5$ kpc is the
height scale.

A condition to constrain the parameters in the previous mass
models is that the total local projected mass density within a
distance of $(0.3-1.1)$ kpc of the galactic plane is in the range
$(40-85)~M_{\odot}$ pc$^{-2}$. \cite{bss} In addition, we require
that the local value of the rotation curve is $v_{\rm
rot}(R_0)=\sqrt{v_{\rm b}^2+v_{\rm d}^2+v_{\rm h}^2+v_{\rm
wd}^2}\simeq (220\pm 20)$ km s$^{-1}$ and that at the LMC distance
is $v_{\rm rot}(LMC)\simeq (240\pm 40)$ km s$^{-1}$.

\section{Microlensing optical depth, rate and event duration}

When a MACHO of mass $m_{\rm i}$ (the i suffix refers to the
considered lens population, i.e., thin and thick stellar disk,
white dwarfs and halo MACHOs) is sufficiently close to the line of
sight between us and a star in the LMC, the light from the source
suffers a gravitational deflection and the original star
brightness increases by \cite{pc86}
\begin{equation}
A=\frac{u^2+2}{u(u^2+4)^{1/2}}~ . \label{eq:bb}
\end{equation}
Here $u=d/R_{\rm E}$ ($d$ is the distance of the MACHO from the
line of sight) and $R_{\rm E}$ is the Einstein radius defined as:
\begin{equation}
R_{\rm E}^2=\frac{4Gm_{\rm i}D_{\rm S}}{c^2}x(1-x)~,
\end{equation}
with $x=D_{\rm L}/D_{\rm S}$, where $D_{\rm S}$ is the
source-observer distance and $D_{\rm L}$ is the distance to the
lens.

The microlensing optical depth is defined as
\begin{equation}
\tau_{\rm i}=\int_0^{D_{\rm S}} n_{\rm i}(D_{\rm L})\pi R_{\rm
E}^2~{\rm d} D_{\rm L} = \frac{4\pi GD_{\rm S}^2}{c^2}\int_0^1
\rho_{\rm i}(x) x (1-x)~{\rm d}x~,
\end{equation}
where  $\rho_{\rm i}(x)=m_{\rm i} n_{\rm i}(x)$ is the lens mass
density for the i-th lens object galactic population, so that the
galactic total optical depth is $\tau_{\rm tot}=\sum_{\rm i}
\tau_{\rm i}$. Of course, the total optical depth should also
include the contribution from self-lensing $\tau_{\rm self}$ by
lenses in the LMC.

It follows that the mean position of the i-th lens object
population is defined as:
\begin{equation}
<x_{\rm i}>=\frac{1}{\tau_{\rm i}}\int \left(\frac{{\rm
d}\tau_{\rm i}}{{\rm d}x}\right)~x~{\rm d}x~. \label {xi}
\end{equation}

The microlensing rate $\Gamma_{\rm i}$ (i.e. the number of events
per unit time and per monitored star) due to the i-th lens object
population is given by \cite{griest}
\begin{equation}
\Gamma_{\rm i}=2\int R_{\rm E}(x) n_{\rm i}(x) v_{\rm L} f({\bf
v}_{\rm L} - {\bf v}_{\rm t}) f({\bf v}_{\rm S}) ~{\rm d}x ~{\rm
d}{\bf v}_{\rm L} ~{\rm d}{\bf v}_{\rm S}, \label{gammai}
\end{equation}
where ${\bf v}_{\rm L}$, ${\bf v}_{\rm S}$ and ${\bf v}_{\rm t}$
are the lens, source and microlensing tube two-velocities in the
plane transverse to the line of sight. Therefore, the integral
above is actually a five-dimensional integration.

As usual, we calculate the transverse tube velocity by
\begin{equation}
v_{\rm t}^2 (x) = (1-x)^2 v_{\odot}^2 + x^2 v_{\rm S}^2 +
2x(1-x)v_{\odot}v_{\rm S} \cos\theta~,
\end{equation}
where ${\bf v}_{\odot}$ is the local velocity transverse to the
line of sight and $\theta$ is the angle between ${\bf v}_{\odot}$
and ${\bf v}_{\rm S}$.

In the rate calculation, eq. (\ref{gammai}), we consider, as
usual, a microlensing amplification threshold $A_{\rm th}=1.34$,
corresponding to $u_{\rm th}=1$. Moreover, the velocity
distribution functions $f({\bf v}_{\rm L} - {\bf v}_{\rm t})$ and
$f({\bf v}_{\rm S})$ are assumed to have a Maxwellian form with
one-dimensional dispersion velocity $\sigma_{\rm i}$ different for
each lens and source populations. We take $\sigma_{\rm d}=30$ km
$s^{-1}$ for the stellar disks and $\sigma_{\rm h}=156$ km
$s^{-1}$ for the bulge and halo populations. In the case of white
dwarfs distributed according to the spheroidal model we take
$\sigma_{\rm wd}=100$ km $s^{-1}$, while for the disk shape
distribution we consider $\sigma_{\rm wd}=50$ km $s^{-1}$.
\cite{jms}

From equation (\ref{gammai}), the differential rate distribution
in event duration $t_{\rm E}$ (defined as the time taken by the
lens to travel the distance $R_{\rm E}$ in the direction
orthogonal to the line of sight) is
\begin{equation}
\frac{{\rm d}\Gamma_{\rm i}}{{\rm d}t_{\rm E}}=K\int_0^1 ~{\rm
d}x~ \rho_{\rm i}(x) ~y^4 e^{-y^2} \int_0^{2\pi} ~{\rm d}\theta
\int_0^{\infty}~{\rm d}z~ z~ e^{-z^2-\eta^2} \int_0^{2\pi}~{\rm
d}\beta ~e^{2\eta y\cos\beta}~, \label{5dim}
\end{equation}
where $K=-D_{\rm S}\sigma_{\rm L}^2/(\pi^2 m_{\rm L})$, $y=v_{\rm
L}/\sigma_{\rm L}$, $z = v_{\rm S}/\sigma_{\rm S}$, $\eta=v_{\rm
t}/\sigma_{\rm L}$ and $\theta$, $\beta$ are the angles between
${\bf v}_{\rm S}$, ${\bf v}_{\rm L}$ and ${\bf v}_{\odot}$. By
using $t_{\rm E}=R_{\rm E}/|{\bf v}_{\rm L} - {\bf v}_{\rm t}|$,
it follows that
\begin{equation}
y=2\frac{\sqrt{4G m_{\rm L} D_{\rm S} x (1-x)}}{c\sigma_{\rm L}
t_{\rm E}}~. \label{y}
\end{equation}
\begin{figure}
      \begin{minipage}[b]{8.5cm}
          {\includegraphics[width=12cm]{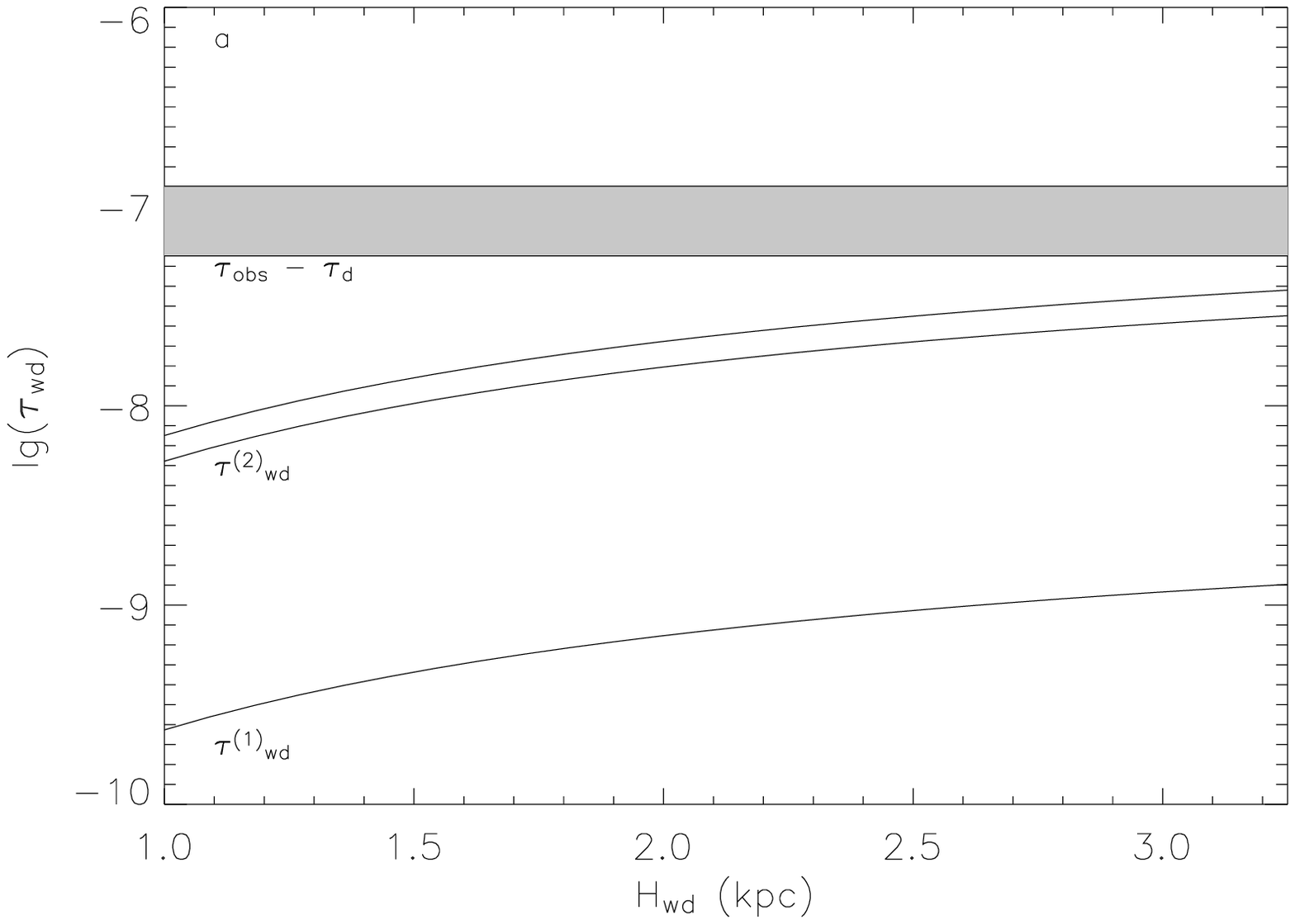}}
      \end{minipage}
      \begin{minipage}[b]{8.5cm}
          {\includegraphics[width=12cm]{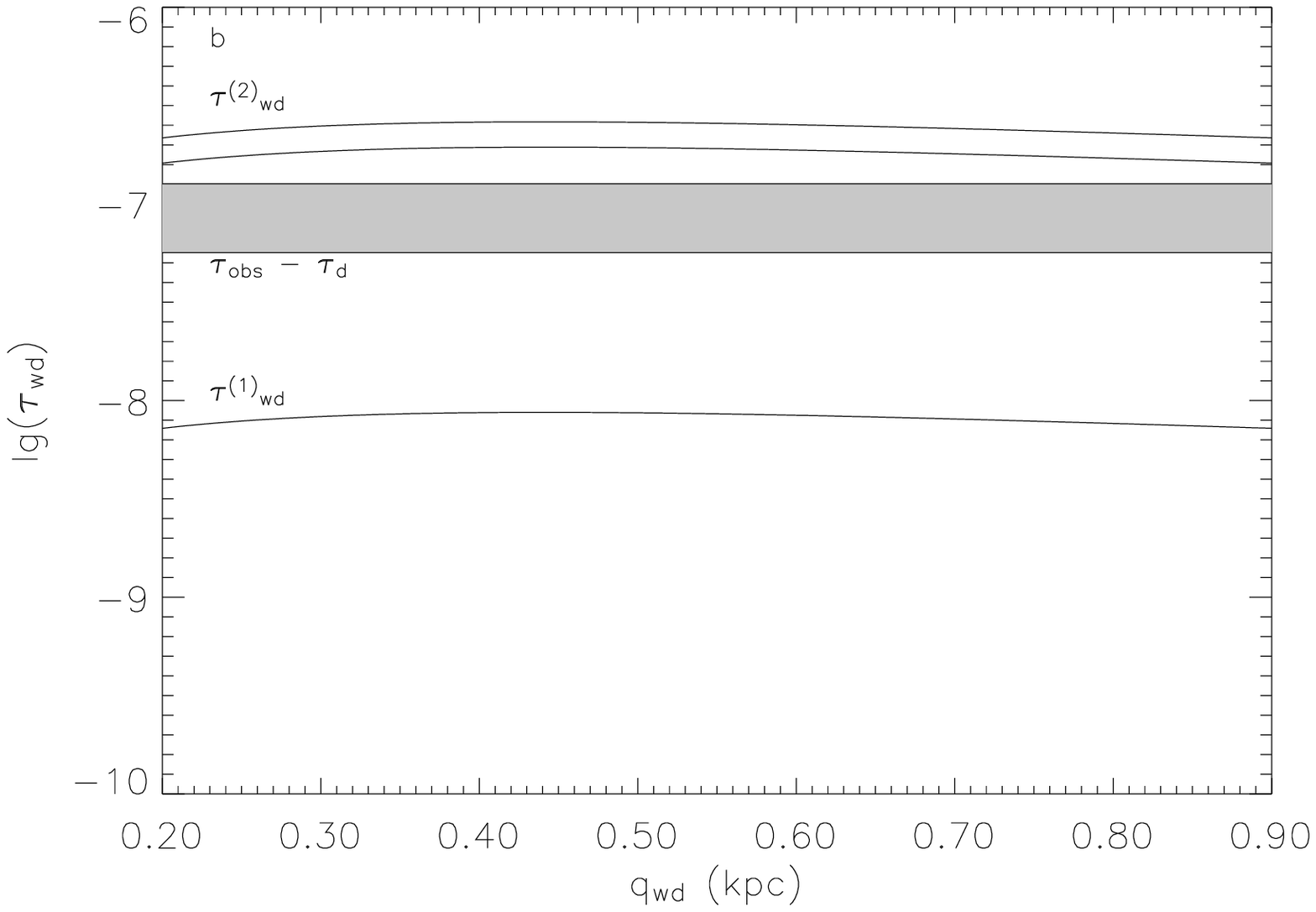}}
      \end{minipage}
  \caption{In panel a the optical depth for the two extremal white dwarf
disk models is shown as a function of the white dwarf vertical
scale height $H_{\rm wd}$. For comparison, the grey band shows the
difference between the observed optical depth
($\tau=1.2^{+0.4}_{-0.3}\times 10^{-7}$) and that estimated for
the standard stellar disk. The curves labelled as $\tau_{\rm
wd}^{(2)}$ and $\tau_{\rm wd}^{(1)}$ correspond to $\rho_{\rm
wd}(R_0)=3.4^{+0.5}_{-0.5}\times 10^{-3}~M_{\odot}$ pc$^{-3}$ and
$\rho_{\rm wd}(R_0)=1.3\times 10^{-4}~M_{\odot}$ pc$^{-3}$,
respectively.  The optical depth for the two extremal white dwarf
spheroidal models is shown in panel b as a function of the white
dwarf flatness parameter $q_{\rm wd}$. Here, we have verified that
the microlensing optical depth does not appreciably depend on the
spheroidal core radius so that we have set $a_{\rm wd}\simeq 2$
kpc. The other components correspond to those of panel a.}
\label{tau}
\end{figure}
The average event duration $<t_{\rm E}>_{\rm i}$ of microlensing
events due to a lens of the i-th population can be computed as
\begin{equation}
<t_{\rm E}>_{\rm i}=\frac{\int_0^{\infty} t_{\rm E}
\epsilon(t_{\rm E})~{\rm d}\Gamma_{\rm i} (t_{\rm
E})}{\int_0^{\infty} \epsilon(t_{\rm E})~{\rm d}\Gamma_{\rm i}
(t_{\rm E})}~, \label {ti}
\end{equation}
while the number of expected microlensing events during an
observation time $t_{\rm obs}$ is given by
\begin{equation}
N_{\rm ev,~i}=N_* t_{\rm obs}\int_0^{\infty}\epsilon (t_{\rm E})~
{\rm d}\Gamma_{\rm i} (t_{\rm E})~, \label{nev}
\end{equation}
where $\epsilon(t_{\rm E})$ is the experimental detection
efficiency towards the LMC given by the MACHO Collaboration
\cite{alcock2001} and $N_*$ is the number of monitored stars.

\section{Model results}

Here we present the results we obtain for the optical depth, the
microlensing rate and the average microlensing event time duration
towards the LMC ($l=280.5^0$, $b=-32.9^0$, $D_{\rm S}=50$ kpc) for
the different mass components of the assumed galactic models.

In Fig. \ref{tau}a we present the optical depth as a function of
the white dwarf vertical height scale $H_{\rm wd}$ for the two
extremal white dwarf disk models with local mass density
$\rho_{\rm wd}(R_0)=3.4^{+0.5}_{-0.5}\times 10^{-3}~M_{\odot}$
pc$^{-3}$ and $1.3\times 10^{-4}~M_{\odot}$ pc$^{-3}$,
respectively. For each value of $\rho_{\rm wd}(R_0)$ and $H_{\rm
wd}$, the white dwarf local projected mass density is determined -
being $\Sigma_{\rm wd}(R_0)=\rho_{\rm wd}(R_0)H_{\rm wd}$ - in
such a way that the local galactic rotation speed, including the
contribution of the galactic halo dark matter $\rho_{\rm
h}(R_0)=[7.9\times 10^{-3}~M_{\odot}$ pc$^{-3}$ $-~\rho_{\rm
wd}(R_0)]$, is in agreement with the experimental constraints on
the galactic rotation curve $v_{\rm rot}(R_0)=220\pm 20$ km
s$^{-1}$ and $v_{\rm rot}(LMC)=240\pm 40$ km s$^{-1}$. For
comparison, we also show in Fig. \ref{tau} (grey band) the
difference between the observed optical depth (estimated by
gravitational microlensing experiments) and that estimated for a
standard stellar disk. \cite{alcock2000} In Fig. \ref{tau}b we
present the optical depth for white dwarf spheroidal models with
local mass density $\rho_{\rm wd}(R_0)=3.4^{+0.5}_{-0.5}\times
10^{-3}~M_{\odot}$ pc$^{-3}$ and $1.3\times 10^{-4}~M_{\odot}$
pc$^{-3}$ as a function of the flatness parameter $q_{\rm wd}$
assuming $a_{\rm wd}=2$ kpc. As one can see, the models
corresponding to the higher values of the local mass density are
clearly excluded by present observations, irrespectively of the
assumed spheroidal core radius $a_{\rm wd}$ (we have verified that
there is only a very weak dependence of $\tau_{\rm wd}$ on $a_{\rm
wd}$) and flatness $q_{\rm wd}$ values. The maximum allowed value
for the local white dwarf mass density of such a distribution can
be estimated by subtracting from the observed value of $\tau$ the
self-lensing and stellar disk contributions \footnote{Several
authors have estimated the self-lensing $\tau_{\rm self}$
contribution, which results to be strongly depend on the assumed
mass density distribution within the LMC. The best estimate to
date is $\tau_{\rm self}=2.6-6.8\times 10^{-8}$. \cite{jms}}.
\cite{jms} The outcome is $\rho_{\rm wd}(R_0)\simeq 1.6\times
10^{-3}~M_{\odot}$ pc$^{-3}$. In the case of white dwarfs in the
disk distribution (see Fig. \ref{tau}a) the observational
constraint for the optical depth is not violated, even for the
maximum value of the local white dwarf mass density.

\begin{figure}[htbp]
\vspace{11.cm} \includegraphics{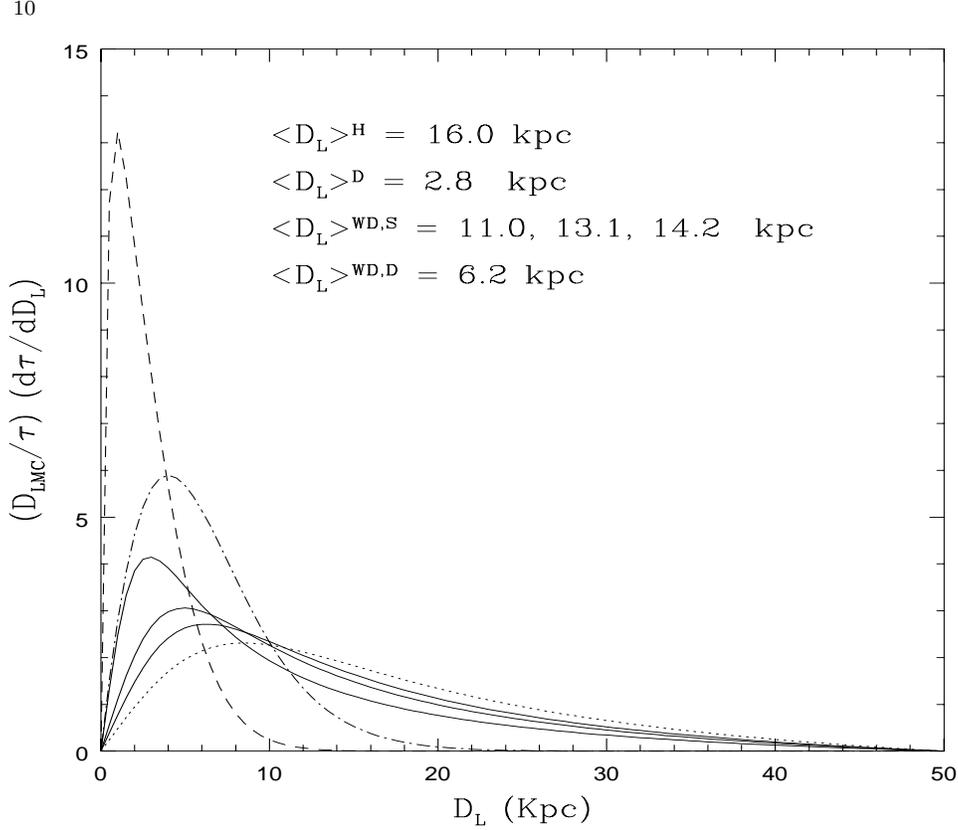} \caption{The differential optical depth
${\rm d}\tau/{\rm d}D_{\rm L}$ as a  function of the observer-lens
distance $D_{\rm L}$ is given for the various galactic components:
stellar disk (dashed line), three white dwarf spheroids
(continuous lines) with $q_{\rm wd}=0.2,~0.4, ~0.6$ (from left to
right),  white dwarf disk (dotted-dashed line) and halo (dotted
line). The average distance $<D_{\rm L}>$ - defined by eq.
(\ref{xi}) - for each lens component is also reported.}
\label{dtau}
\end{figure}

In Fig. \ref{dtau}, the differential optical depth ${\rm
d}\tau/{\rm d}D_{\rm L}$ as a function of the lens distance
$D_{\rm L}$ is given for the various galactic lens component (note
that the results do not depend to the assumed local mass density):
stellar disk (dashed line), white dwarfs in the spheroidal model
(continuous lines) with $q_{\rm wd}=0.2,~0.4, ~0.6$ (from left to
right), white dwarfs in the disk distribution (dotted-dashed line)
for $H_{\rm wd}=3$ kpc, and dark halo (dotted line). In the same
Figure we also give the mean distance $<D_{\rm L}>$ for each lens
object population. As it is expected, the mean distance increases
from disk to halo populations. However, the large uncertainties
(which result to be of the order of $65\%$ of the mean values)
show how it is difficult to disentangle lens of different
populations by only using considerations on the optical depth.

\begin{figure}
      \begin{minipage}[b]{8.5cm}
          {\includegraphics[width=12cm]{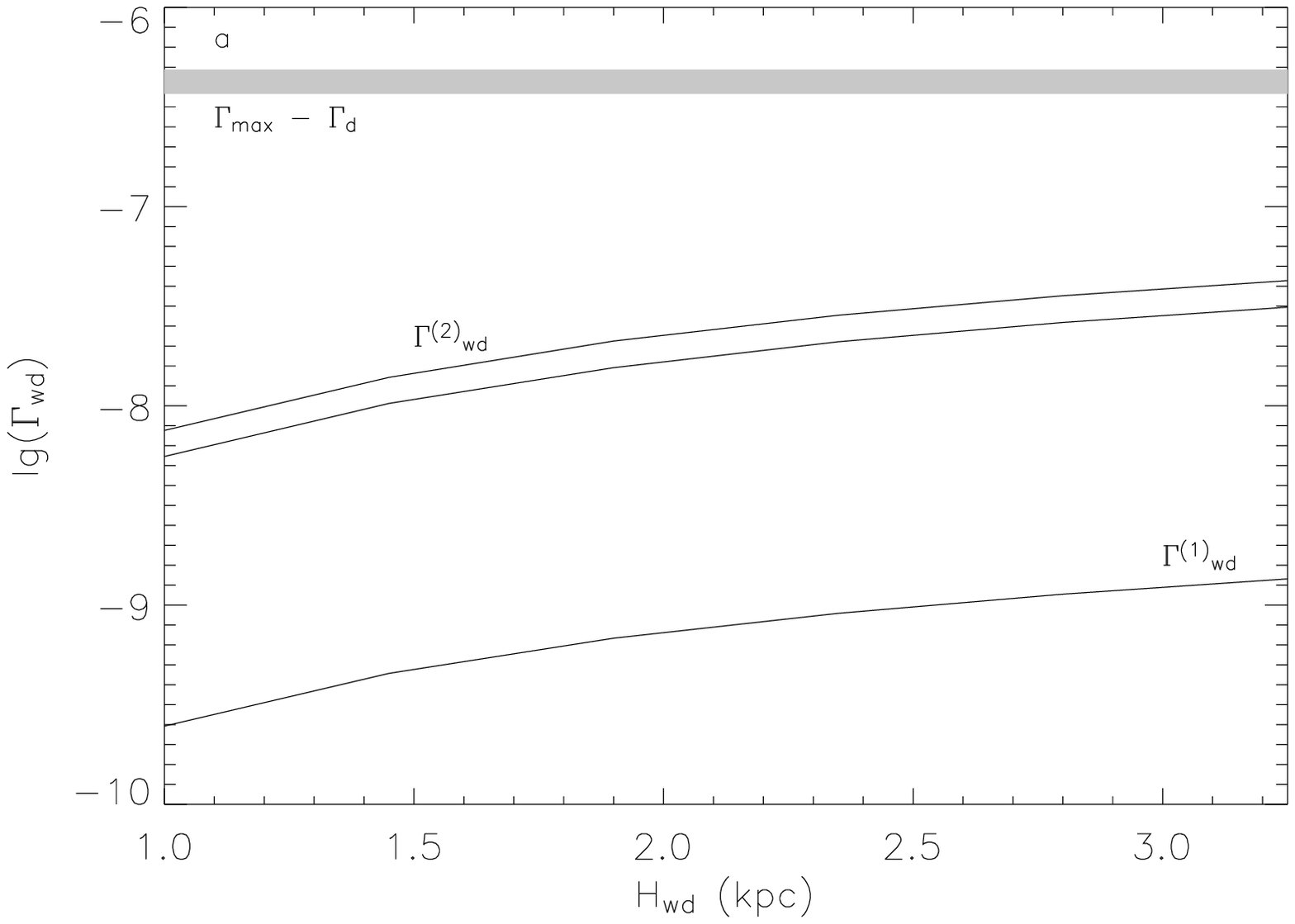}}
      \end{minipage}
      \begin{minipage}[b]{8.5cm}
          {\includegraphics[width=12cm]{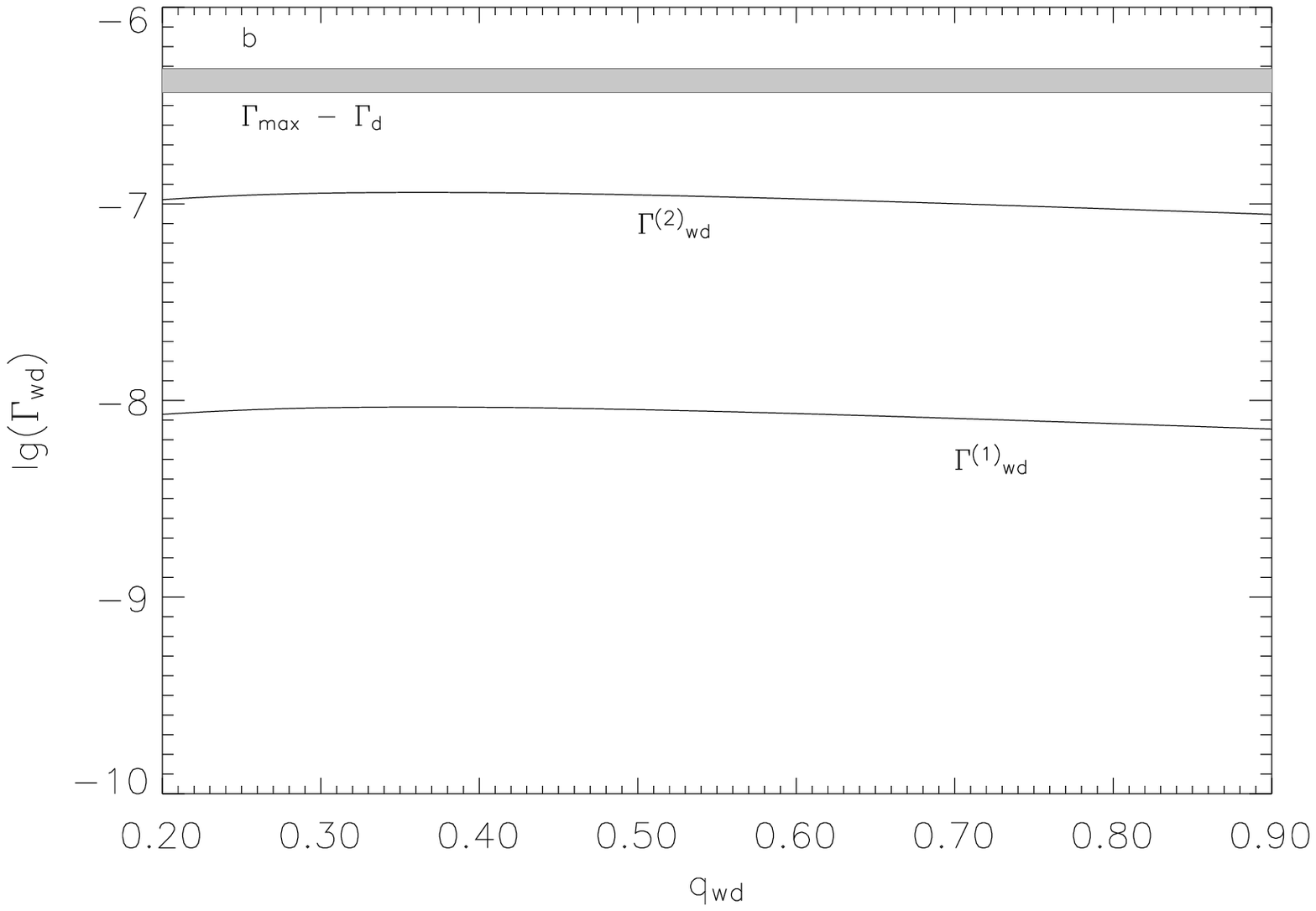}}
      \end{minipage}
 \caption{In panel a, the microlensing rate $\Gamma_{\rm wd}$
expected for the white dwarf disk model is shown as a function of
the disk scale height $H_{\rm wd}$. The curve labelled
$\Gamma^{(1)}_{\rm wd}$ corresponds to a local white dwarf mass
density $\rho_{\rm wd}(R_0)=1.3\times 10^{-4}~M_{\odot}$ pc$^{-3}$
while the band labelled $\Gamma^{(2)}_{\rm wd}$ corresponds to
mass density values $\rho_{\rm wd}(R_0)=3.4^{+0.5}_{-0.5}\times
10^{-3}~M_{\odot}$ pc$^{-3}$. In panel b white dwarfs are assumed
to be distributed according to a spheroidal model with flatness
parameter $q_{\rm wd}$. The curve labelled $\Gamma_{\rm wd}^{(1)}$
corresponds to the same white dwarf mass density as in panel a
while the curve $\Gamma_{\rm wd}^{(2)}$ corresponds to $\rho_{\rm
wd}(R_0)=1.6\times 10^{-3}~M_{\odot}$ pc$^{-3}$ which follows from
the constraints of the observed optical depth (see Fig. \ref{tau}b
and discussion in the text).} \label{gamma}
\end{figure}
In Fig. \ref{gamma} we present the obtained microlensing rates for
white dwarfs models in the disk distribution (Fig. \ref{gamma}a)
and in the spheroidal distribution (Fig. \ref{gamma}b). The curve
labeled as $\Gamma^{(2)}_{\rm wd}$  in Fig. \ref{gamma}a is given
for $\rho_{\rm wd}(R_0)=3.4^{+0.5}_{-0.5}\times 10^{-3}~M_{\odot}$
pc$^{-3}$ while in Fig. \ref{gamma}b we have adopted only the
upper limit for the white dwarf local density allowed by
observations (see the discussion above). For comparison, we also
give the differnce (grey band) between the maximum allowed rate
$\Gamma_{\rm max}=(3.8-5)\times 10^{-7}$ yr$^{-1}$ (which
corresponds to the 13-17 events detected so far in microlensing
experiments towards the LMC, \cite{alcock2000}) and that estimated
for the standard stellar disk. To estimate $\Gamma_{\rm max}$ we
have assumed a total number of observed LMC stars $N_*=1.19\times
10^7$ during an observation time of $t_{\rm obs}=5.7$ yr and
adopted the instrument detection efficiency $\epsilon (t_{\rm E})$
given by the MACHO Collaboration. \cite{alcock2001}  Notice that
we are neglecting the source spatial distribution. The only effect
of this is that we are unable to estimate the expected event
number as a function of the direction and the number of expected
self-lensing events.

Once the microlensing rate for each lens population has been
calculated, it is straightforward to obtain the expected number of
microlensing events by eq. (\ref{nev}).
\begin{table}[htbp]
\ttbl{30pc}{The expected number of events for different types of
lenses and models are reported. For white dwarfs in the spheroidal
models (first four lines) we assume $a_{\rm wd}=2~{\rm kpc}$. The
assumed mass values are $m_{\rm d}=0.4~M_{\odot}$ for disk stars,
$m_{\rm h}=0.5~M_{\odot}$ for halo lenses and $m_{\rm
wd}=0.65~M_{\odot}$ for white dwarfs. We assume that about $20\%$
of the halo dark matter is made of MACHOs. The models labelled
with $S$ correspond to white dwarfs distributed in spheroidal
models with flatness parameter $q_{\rm wd}$ given as subscript.
The models labelled as $D$ corresponds to white dwarfs distributed
in disk models with height scale appearing as subscript (in kpc).
The superscript $l$ corresponds to white dwarf local mass-density
$\rho_{\rm wd}(R_0)=1.3\times 10^{-4}~M_{\odot}$ pc$^{-3}$ while
$h$ corresponds to $\rho_{\rm wd}(R_0)=1.6\times
10^{-3}~M_{\odot}$ pc$^{-3}$ for spheroidal models and to
$\rho_{\rm wd}(R_0)=4.4\times 10^{-3}~M_{\odot}$ pc$^{-3}$ for
disk models (for details see text).}
{\begin{tabular}{|c|c|c|c||c|c|c|}\\
\multicolumn{4}{c}{} \\[6pt]\hline
WD model & $N_{\rm d}$ & $N_{\rm h}$  & $N_{\rm wd}$ & $<t_{\rm d}>$&$<t_{\rm h}>$&$<t_{\rm wd}>$\\
\hline
\hline $S_{0.2}^l $ & $2.2$ & $8.0$ & $0.4$ & $47\pm 18$ & $50\pm 26$ & $58\pm 25$\\
\hline $S_{0.6}^l$&$2.2$&$8.0$&$0.4$ & $47\pm 18$&$50\pm 26$&$66\pm 27$\\
\hline $S_{0.2}^h$&$2.4$&$7.1$&$4.1$  & $53\pm 21$&$50 \pm 26$&$57\pm 24$ \\
\hline $S_{0.6}^h$&$2.4$&$7.1$&$4.4$ & $53\pm 20 $&$51 \pm 26 $&$66\pm 27$\\
\hline
\hline $D_3^l$&$2.0$&$8.3$&$0.1$ & $50\pm 20$&$ 31 \pm 14$&$45\pm 14$\\
\hline $D_5^l$&$2.0$&$8.4$&$0.1$ & $50\pm 20$&$31\pm 14$&$42\pm 14$ \\
\hline $D_1^h$&$2.0$&$8.4$&$0.3$ & $49\pm 19$&$ 31\pm 14$&$ 58\pm 22  $\\
\hline $D_3^h$&$2.0$&$8.4$&$1.2$ & $49\pm 19$&$31\pm 14$&$ 42 \pm 13$\\
\hline
\end{tabular}}
\label{Tablenev}
\end{table}

The obtained results are given in Table \ref{Tablenev} for some
selected models. As one can see, the expected number of events for
white dwarfs in the spheroidal model results in the range
$0.4-4.4$, for the lower ($\rho_{\rm wd}(R_0)=1.3\times
10^{-4}~M_{\odot}$ pc$^{-3}$ given by \cite{oppenheimer}) and
upper ($\rho_{\rm wd}(R_0)=4.4\times 10^{-3}~M_{\odot}$ pc$^{-3}$
taking two standard deviations for the local white dwarf mass
density estimated by \cite{holberg}) local mass density values,
respectively. For white dwarfs in the disk model the expected
number of events is $0.1-1.2$, for the lower and upper local mass
density values, respectively. Clearly, if one considers that the
total number of events observed by the MACHO Collaboration in a
5.7 years campaign towards the LMC is $13-17$ \cite{alcock2000}
and that 3-4 of these events are expected to be due to
self-lensing \cite{jms}, the obtained results may be considered as
consistent with observations for all white dwarf models.

\begin{figure}[htbp]
\vspace{11.0cm} \includegraphics{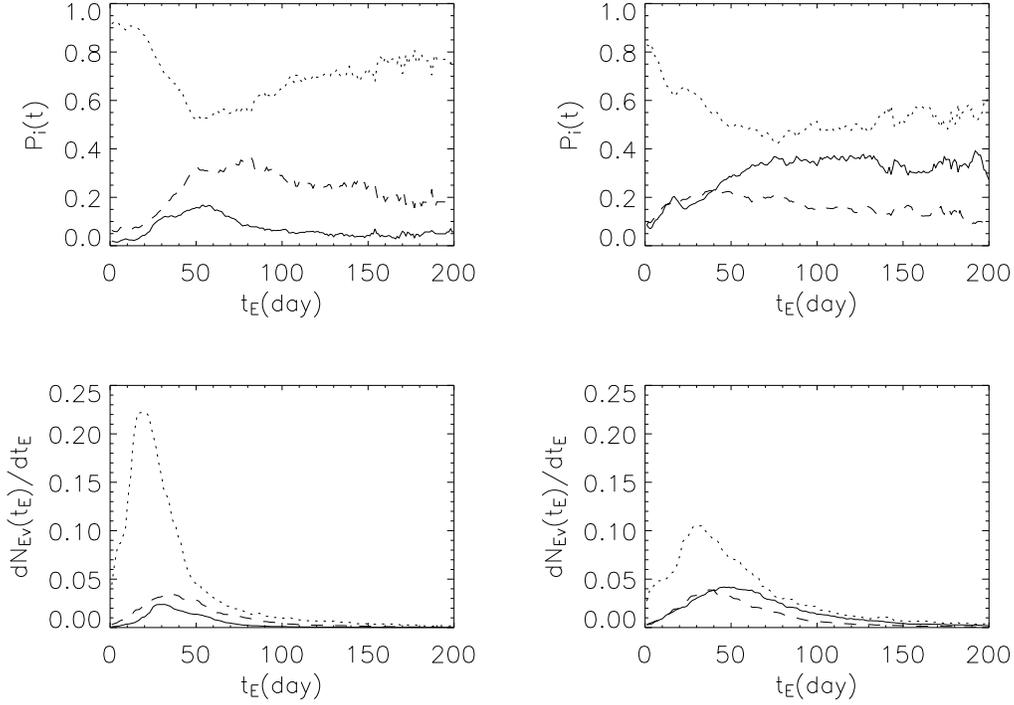} \caption{The differential event
number ${\rm d}N_{\rm ev}(t_{\rm E})/{\rm d}t_{\rm E}$ (lower
panels) is given as a function of the duration $t_{\rm E}$, for
the fourth (right) and eighth (left) model in the Table. For each
lens component, the curves are normalized to the respective total
number of events quoted in the Table. In the upper panels, for the
same models, the probability for each lens population is also
given as a function of $t_{\rm E}$. As in the previous Figures, we
adopt dotted, dashed and continuous lines for halo, disk and white
dwarf lens components, respectively. We note that the fluctuations
in the curves are due to the Montecarlo method.} \label{figeventi}
\end{figure}

In principle, the mean time duration of the microlensing events
caused by each of the three lens components could be a better
indicator to investigate the presence of a white dwarf component
through microlensing searches. In the same Table we give the mean
event duration $<t_{\rm E}>$, with the relative uncertainty
$\sigma_{<t_{\rm E}>}$, of microlensing events from stellar disk
lenses, halo lenses and selected models of white dwarf
distributions. To estimate $\sigma_{<t_{\rm E}>}$ we use a
Montecarlo program \footnote{In particular, we count the number of
events falling in each duration bean ($t,~t+\delta t$) by
performing a five-dimensional integration on $x, ~{\bf v}_{\rm
S},~{\bf v}_{\rm L}$ (see eq. \ref{5dim}). For each choice of the
five parameters above made by the integration (through Montecarlo
methods) routine, the event duration is uniquely determined so
that one can calculate the number of occurrence of a given
duration value.} to obtain the event duration distributions (which
are given in Fig. \ref{figeventi}) and we define $\sigma_{<t_{\rm
E}>}$ so that $<t_{\rm E}>-\sigma_{<t_{\rm E}>}$, $<t_{\rm
E}>+\sigma_{<t_{\rm E}>}$ is the duration range containing $50\%$
of expected events.

As one can see from the Table, it is extremely difficult to really
distinguish between the different lens populations since they have
comparable average duration values. This is also more clear in
Fig. \ref{figeventi} in which the probability distribution (upper
panels) and the distribution of the number of events (lower
panels) for each galactic population are shown as a function of
the microlensing duration for the fourth and eighth model given in
the Table. The curves shown allow to directly estimate the
probability that an observed event belong to one of the considered
populations.

\section{Conclusions}
In the present paper we have considered, in addition to the
standard stellar disk, bulge and halo lens populations,
gravitational microlensing towards the LMC also from a population
of white dwarfs distributed according to two different models:
thick disk and spheroidal shape with various flatness values.

By comparing optical depth theoretical results and observational
data towards the LMC we find for the local mass density of white
dwarfs in the spheroidal distribution an upper limit of about
$\rho_{\rm wd}(R_0)\simeq 1.6\times 10^{-3}~M_{\odot}$ pc$^{-3}$.
For white dwarfs distributed according to disk models, instead,
available observations are unable to set an upper limit on the
white dwarf local mass density value.

We have then calculated the expected microlensing rate from each
lens population model, the average event duration $<t_{\rm
E}>_{\rm i}$ and the expected number of events $N_{\rm ev, i}$ for
each galactic lens population by taking into account the detection
efficiency given by the MACHO Collaboration. \cite{alcock2001}

In the Table our main results for selected lens models are
presented. As one can see, the expected number of events for white
dwarfs in the spheroidal model results in the range $0.4-4.4$, for
the lower and upper local mass density values, respectively. For
white dwarfs in the disk model the expected number of events is
$0.1-1.2$, for the lower and upper local mass density values.
Clearly, if one considers that the total number of observed events
is $13-17$ \cite{alcock2000} and that $3-4$ events can be
accounted for by self-lensing, \cite{jms} the obtained results may
be considered as consistent with observations. The event durations
for the various adopted models are given in the Table. From this
Table and Fig. \ref{figeventi} it is clear how it is difficult to
distinguish between the various lens populations, since the
obtained distributions look similar. Therefore, microlensing
observations alone cannot allow at present to understand the
nature of the lens.  Indeed, in order to distinguish among
different lens populations, one should eliminate the degeneracy in
microlensing observations (which give as a result the event
duration and the source amplification) since a direct measurement
of the MACHO distance, mass or transverse velocity is generally
not possible. The best way to solve the issue is to perform
parallax observations of microlensing events since it allows a
direct measurement of the MACHO distance to the observer.
\cite{gg,hk,soszy} On the other hand, this requires telescopes in
orbit around the Sun, which is probably rather far in the future.

In any case, present and future microlensing and pixel lensing
observations towards different directions - e.g. M31 \cite{calchi}
or other nearby galaxies - should allow to increase the available
statistics of microlensing events towards several directions
getting, through an accurate data analysis, more information about
the possible presence of different lens populations in the Galaxy.
Moreover, it will be possible to find an increasing number of
microlensing events for which parallax measurements (or a direct
imaging of the lens through the next generation of space-based
telescopes) may allow to solve the parameter degeneracy allowing a
direct estimate of the lens mass and distance.

\end{document}